\documentclass[12pt]{elsart}
\usepackage{amstex}
\usepackage{amssymb}
\usepackage{citesort}
\usepackage{epsfig}

\begin{document}

\begin{flushright}
BARI-TH 218/95 
\end{flushright}

\begin{frontmatter}

\title{The Lattice Schr\"odinger Functional and the\\ Background Field
Effective Action}

\author[INFN,Dep]{P. Cea\thanksref{emcea}},
\author[INFN]{L. Cosmai\thanksref{emcosmai}} and
\author[INFN,Dep]{A. D. Polosa\thanksref{empolosa}}
\address[INFN]{INFN - Sezione di Bari - 
Via Amendola, 173 - I 70126 Bari - Italy} 
\address[Dep]{Dipartimento di Fisica Univ. Bari -
Via Amendola, 173 - I 70126 Bari - Italy} 
\thanks[emcea]{E-mail: cea@@bari.infn.it}
\thanks[emcosmai]{E-mail: cosmai@@bari.infn.it}
\thanks[empolosa]{E-mail: polosa@@bari.infn.it}

\begin{abstract} 
We propose a new method that by using the lattice Schr\"odinger
functional allows to investigate the effective
action for external background fields in lattice gauge theories. We
show that this method gives sensible results for the case of
four-dimensional U(1) gauge theory in an external constant magnetic
field.
\end{abstract}
\end{frontmatter}

\section{Introduction}

The Euclidean Schr\"odinger functional in Yang-Mills theories without
matter fields is defined by
\begin{equation}
\label{Eq1}
{\mathcal{Z}} \left[ A^{(f)}, A^{(i)} \right] = 
 \langle  A^{(f)}  |  \exp(-HT) {\mathcal{P}} | 
A^{(i)} 
\rangle  \,,
\end{equation}
where the operator ${\mathcal{P}}$ projects onto the physical states
and $H$ is the pure gauge Yang-Mills Hamiltonian in the temporal
gauge.
$A^{a(i)}_k\left( \vec{x} \right)$ and
$A^{a(f)}_k\left( \vec{x} \right)$ are classical gauge fields, and the
state $| A \rangle$ is such that
\begin{equation}
\label{Eq2}
\left \langle A | \Psi \right \rangle = \Psi(A)
\end{equation}                         
for all wave functionals $\Psi(A)$.
From Equation\eqref{Eq1}, inserting an orthonormal basis
$\{|\Psi_n\rangle\}$ of gauge invariant energy eigenstates, it follows
\begin{equation}
\label{Eq3}
{\mathcal{Z}} \left[ A^{(f)}, A^{(i)} \right] = 
\sum_n \exp(-E_n T) \Psi_n\left(A^{(f)}\right)
\Psi^{*}_n\left(A^{(i)}\right) \,,
\end{equation}
where $E_n$ are the energy eigenvalues.
Note that Eq.\eqref{Eq3} implies that 
${\mathcal{Z}} \left[ A^{(f)}, A^{(i)} \right]$ is invariant under
arbitrary gauge transformations of the fields $A^{(f)}$ and 
$A^{(i)}$. 

The authors of Refs.\cite{Wolff86,Luscher92} studied the Schr\"odinger
functional in lattice gauge theories. It turns out that the
Schr\"odinger functional has a well-defined continuum limit and,
moreover, it is amenable to numerical simulations~\cite{Luscher92}.
The lattice Schr\"odinger functional is given by
\begin{equation}
\label{Eq4}
{\mathcal{Z}} \left[ U^{(f)}, U^{(i)} \right] = 
\int {\mathcal{D}}U \exp(-S)  \,.
\end{equation}
In Equation\eqref{Eq4} the action $S$ is the standard Wilson action
modified to take into account the boundaries at $x_4=0$, and
$x_4=T$~\cite{Luscher92}:
\begin{equation}
\label{Eq5}
S = \frac{1}{g^2} \sum_{x, \mu > \nu} w_{\mu\nu}(x) \text{Tr}
\left[1-U_{\mu\nu}(x) \right] \,,
\end{equation}
where the $w_{\mu\nu}(x)$'s are equal to $1/2$ for the spatial
plaquettes at $x_1=0$ and $x_1=T$, otherwise they are equal to 1.
Obviously, in Eq.\eqref{Eq4} one integrates over the links $U_\mu(x)$
with the fixed boundary values
\begin{equation}
\label{Eq6}
U(x)|_{x_4=0} = U^{(i)} \, ,  \quad  U(x)|_{x_4=T} = U^{(f)} \,.
\end{equation}
The external links $U^{(i)}$ and $U^{(f)}$ are the lattice
implementation of the smooth classical boundary fields $A_k^{a(i)}(x)$
and  $A_k^{a(f)}(x)$. It is worthwhile to stress that the functional
Eq.\eqref{Eq4} is invariant under arbitrary lattice gauge
transformations of the boundary links.  Moreover, we note that in the
numerical simulations of the lattice Sch\"odinger functional one can
assume periodic boundary conditions in the spatial directions, while
the periodicity in the time direction is lost if $U^{(i)} \ne
U^{(f)}$.

The aim of the present paper is to investigate the lattice effective
action for external background fields. Let us consider a static
external background field 
$\vec{A}^{\text{ext}}(\vec{x}) =  \vec{A}_a^{\text{ext}}(\vec{x})
\lambda_a/2$, where $\lambda_a/2$ are the generators of the SU(N) Lie
algebra. On the lattice the dynamical variables are the links
$U_\mu(x)$. The natural relation between the continuum gauge field and
the corresponding lattice link is given by
\begin{equation}
\label{Eq7}
U_\mu^{\text{ext}}(x) = {\text{P}} \exp \left\{ + iag  \int_0^1 dt \,
A_\mu^{\text{ext}}(x+ at {\hat{\mu}}) \right\}
\end{equation}
where ${\text{P}}$ is the path-ordering operator. We can now define
the lattice effective action for the background field
$A_\mu^{\text{ext}}(\vec{x})$ by means of the lattice Schr\"odinger
functional Eq.\eqref{Eq4}as follows:
\begin{equation}
\label{Eq8}
\Gamma\left[ \vec{A}^{\text{ext}} \right] = -\frac{1}{T} 
\ln \left\{ \frac{{\mathcal{Z}}[U^{\text{ext}}]}{{\mathcal{Z}}(0)} \right\}
\end{equation}
where $T$ is the extension in the Euclidean time. In
Equation\eqref{Eq8} we define 
${\mathcal{Z}}[U^{\text{ext}}]=
{\mathcal{Z}}[U^{\text{ext}},U^{\text{ext}}]$, and
${\mathcal{Z}}(0)$ means the lattice Schr\"odinger functional without
external background field ($U_\mu^{\text{ext}} =1$).
From the previous discussion it is clear that 
$\Gamma\left[ \vec{A}^{\text{ext}} \right]$ is invariant for lattice gauge
transformations of the external links $U_\mu^{\text{ext}}(\vec{x})$. In
particular, if we consider background fields that give rise to constant field
strength, then it is easy to show that 
$\Gamma\left[ \vec{A}^{\text{ext}} \right]$ is proportional to the spatial
volume $V$. In this case one is interested in the density of the effective
action:
\begin{equation}
\label{Eq9}
\varepsilon\left[ \vec{A}^{\text{ext}} \right] =
-\frac{1}{V \cdot T} \ln \left[ 
\frac{{\mathcal{Z}}[U^{\text{ext}}]}{{\mathcal{Z}}(0)} \right] \,.
\end{equation}
Note that our definition of the lattice effective action uses the lattice
Schr\"odinger functional with the same boundary fields at $x_4=0$ and
$x_4=T$.
As a consequence we can glue the two hyperplanes $x_0=0$ and $x_0=T$ together.
Thus, we end up in a lattice with periodic conditions in the time direction
too. Therefore we have
\begin{equation}
\label{Eq10}
{\mathcal{Z}}[U^{\text{ext}}]= \int {\mathcal{D}} U \, \exp(-S) \,.
\end{equation}
with the constraints 
\begin{equation}
\label{Eq11}
U_\mu(x)|_{x_4=0} = U_\mu^{\text{ext}} \,.
\end{equation}
In other words the Schr\"odinger functional Eq.\eqref{Eq10} is given by the
standard partition function on a periodic lattice with a cold wall at
$x_4=0$.
Note that, due to the lacking of free boundaries, the action in Eq.\eqref{Eq10}
is now the familiar Wilson action 
\begin{equation}
\label{Eq12}
S=S_W= \frac{1}{g^2} \sum_{x, \mu > \nu} {\text{Tr}} \left[ 1 - U_{\mu\nu}(x)
\right] \,.
\end{equation}
In the remainder of this paper we test our definition of lattice effective
action in the case of the simplest lattice gauge theory, namely the U(1)
compact pure gauge theory without matter fields.

\section{U(1) effective action}

It is known that for Wilson action on a four-dimensional 
lattice with periodic boundary
conditions, the electric charge is confined for $\beta < \beta_c \simeq 1.01$,
while for $\beta > \beta_c$ the gauge system is made of free photons. Moreover,
the phase transition is triggered by the condensation of magnetic
monopoles~\cite{Banks77,DeGrand80,Kerler95,DelDebbio95}. 
In the seminal paper by DeGrand
and Toussaint~\cite{DeGrand80} it has been shown that external magnetic
fields are sensitive to the lattice magnetic monopoles. In particular, it turns
out that for strong coupling (in four dimensions) monopoles screen external
magnetic fields, while for weak coupling the magnetic fields penetrates into
the lattice. Thus it is worthwhile to investigate the lattice effective action
for constant background magnetic fields. In the continuum the vector potential
corresponding to a constant magnetic field along the $x_3$ direction is given
by
\begin{equation}
\label{Eq12a}
A_k^{\text{ext}}(\vec{x}) = \delta_{k,2} x_1 B
\end{equation}
in the Landau gauge. From Eq.\eqref{Eq7} it follows 
\begin{equation}
\label{Eq13}
\begin{split}
U_2^{\text{ext}}(x) = \exp \left[ i a g B x_1 \right] 
= \cos(agBx_1) + i \sin(agBx_1) \,,\\
U^{\text{ext}}_1(x) =  U^{\text{ext}}_3(x) = U^{\text{ext}}_4(x) =  1 \,.
\end{split}
\end{equation}
The periodic boundary conditions result in the quantization of the external
magnetic field
\begin{equation}
\label{Eq14}
a^2 g B = \frac{2 \pi}{L_1} n^{\text{ext}}
\end{equation}
where $n^{\text{ext}}$ is an integer and $L_1$ is the lattice extension in the
$x_1$ direction in lattice units. The action corresponding to the
links~\eqref{Eq13} on a lattice of size $L_1L_2L_3L_4$ with periodic boundary
conditions is readily evaluated:
\begin{equation}
\label{Eq15}
S^{\text{ext}}=\beta \Omega \left[ 1 - \cos(a^2gB) \right]
\end{equation}
where $\beta=1/g^2$, and $\Omega=L_1L_2L_3L_4$ is the lattice volume.
Note that in the naive continuum limit $S^{\text{ext}}$ reduces to
$V\cdot T \,\frac{B^2}{2}$. 
Equation\eqref{Eq15} shows that, in order to be close
to the continuum limit on a finite lattice, we must require that $a^2gB \ll 1$.
This in turn implies that $L_1 \gg 1$. Moreover, in order to select
the ground state contribution in the sum~\eqref{Eq3}, we also need
$L_4 \gg 1$. As a consequence we performed
our numerical simulations on lattices with size $L_1=64$, $L_4=32$ and
$L_2=L_3=6,\,8,\,10$. We use the standard Metropolis algorithm to
update gauge configurations. The links belonging to the time slice
$x_4=0$ are frozen to the configuration~\eqref{Eq13}. In addition we
impose that the links at the spatial boundaries are fixed according
to~\eqref{Eq13}. In the continuum this condition amounts to the usual
requirement that the fluctuations over the background field vanish at
the infinity.

As a preliminary step, it is important to test the behaviour of the
magnetic field. To this end we look at the field strength tensor for a
given time slice. We define
\begin{equation}
\label{Eq16}
F_{\mu\nu}(x_4)= \sqrt{\beta} \, \left\langle \frac{1}{V}
\sum_{\vec{x}} \sin \theta_{\mu\nu} (\vec{x}, x_4) \right\rangle 
\end{equation}
where $\theta_{\mu\nu} (\vec{x}, x_4)$ is the plaquette angle in the
$(\mu,\nu)$ plane. Clearly for $x_4=0$ (or $x_4=L_4$ due to the
periodic boundary conditions) we have 
\begin{equation}
\label{Eq17}
F_{12}(0) \equiv F_{12}^{\text{ext}} = \sqrt{\beta} \sin \left( \frac{2
\pi}{L_1} n^{\text{ext}} \right) \, ,
\end{equation}
the other components of the field strength tensor being equal to zero.
In Figures~1 and 2 we display
$F_{\mu\nu}$ versus the Euclidean time $x_4$ for the $64 \times 10^2\times
32$ lattice and $n^{\text{ext}}=2$.
As expected, we find that only the component $F_{12}$ of the
field strength tensor is present in our data. 
For $\beta <1$ the external
magnetic field is shielded after a small penetration (Fig.~1). On the other
hand, for $\beta > 1$ (Fig.~2) the field penetrates indicating that the gauge
system supports a long range magnetic field. 

We now turn on the
evaluation of the density of the effective action~\eqref{Eq9}. We face
with the problem of computing a partition function which is the
exponential of an extensive quantity~\cite{Hasenfratz90}. To avoid
this problem we consider the derivative of
$\varepsilon[\vec{A}^{\text{ext}}]$ with respect to $\beta$.
We get
\begin{equation}
\label{Eq18}
\varepsilon^{\prime} \left[ \vec{A}^{\text{ext}} \right] = 
\frac{\partial \varepsilon \left[ \vec{A}^{\text{ext}}
\right]}{\partial \beta} = 
- \frac{1}{\Omega} \,
\left[
\frac{1}{Z[U^{\text{ext}}]} \frac{\partial Z[U^{\text{ext}}]}{\partial
\beta} - \frac{1}{Z[0]} 
\frac{\partial Z[U^{\text{ext}}]}{\partial\beta} \right] \,.
\end{equation}
A straightforward calculation gives:
\begin{equation}
\label{Eq19}
\varepsilon^{\prime} \left[ \vec{A}^{\text{ext}} \right] = 
\left\langle \frac{1}{\Omega} \sum_{x,\mu>\nu} \cos \theta_{\mu\nu}(x)
\right\rangle_0 -
\left\langle \frac{1}{\Omega} \sum_{x,\mu>\nu} \cos \theta_{\mu\nu}(x)
\right\rangle_{A^{\text{ext}}}
\end{equation}
where the subscripts on the average indicate the value of the external
links at the boundaries. As we have already discussed, in the
deconfined region of our gauge system only the magnetic field directed
along the third direction is present. Thus we expect that the main
contribution to the effective action density comes from the plaquettes
in the 1-2 planes. To check this we look at the contributions due to
the plaquettes in the $(\mu,\nu)$-planes to the derivative of the
effective action density for a given time slice:
\begin{equation}
\label{Eq20}
\varepsilon^{\prime}_{\mu\nu} [x_4] = 
\left\langle \frac{1}{V} \sum_{\vec{x}} \cos
\theta_{\mu \nu}(\vec{x},x_4)
\right\rangle_0 -
\left\langle \frac{1}{V} \sum_{\vec{x}} \cos
\theta_{\mu\nu}(\vec{x},x_4)
\right\rangle_{\vec{A}^{\text{ext}}}  \,.
\end{equation}
Figure~3, where we display $\varepsilon^{\prime}_{\mu\nu} [x_4]$
versus $x_4$ for $\beta = 1.1$, is in full agreement with our
expectations.

In Figure~4 we show $\varepsilon^{\prime}[\vec{A}^{\text{ext}}]$
versus $\beta$ for three different lattice sizes and
$n^{\text{ext}}=2$. A few comments are in order. For small $\beta$, 
$\varepsilon^{\prime}[\vec{A}^{\text{ext}}]$ reduces to a constant
value that is the contribution due to the frozen boundaries. It
should be emphasized that the density of the effective action can be
recovered by integrating $\varepsilon^{\prime}$ over $\beta$:
\begin{equation}
\label{Eq21}
\varepsilon[ \vec{A}^{\text{ext}}, \beta] = \int_0^{\beta} \,
d\beta^{\prime} \, 
\varepsilon^{\prime}[ \vec{A}^{\text{ext}}, \beta^{\prime}] \,.
\end{equation}
Obviously, one should subtract in Eq.~\eqref{Eq21} the contribution due
to the boundaries. 

Fig.~4 shows that, in the weak coupling region $\beta \gg 1$, 
$\varepsilon^{\prime}[ \vec{A}^{\text{ext}}]$ tends to the
derivative of the external action Eq.~\eqref{Eq15}
\begin{equation}
\label{Eq22}
\varepsilon^{\prime}_{\text{ext}} = \frac{\partial}{\partial \beta}
\frac{1}{\Omega} S^{\text{ext}} = 1- \cos \left( \frac{2 \pi}{L_1}
n^{\text{ext}} \right) \,.
\end{equation}
This means that for large $\beta$ the effective action agrees with the
classical action 
\begin{equation}
\label{Eq23}
\lim_{\beta \rightarrow \infty} \varepsilon [\vec{A}^{\text{ext}}]
=
\varepsilon_{\text{ext}} [\vec{A}^{\text{ext}}] = \beta
\left[ 1 - \cos \left( \frac{2 \pi}{L_1} n^{\text{ext}} \right)
\right] \,.
\end{equation}
Note that in the continuum limit $a \rightarrow 0$ and $B$ fixed,
Eq.~\eqref{Eq23} gives the classical energy density $B^2/2$. A
remarkable feature of Fig.~4 is the peak near $\beta=1$. In Figure~5
we present the derivative of the effective action for values of
$\beta$ near the critical region $\beta \approx 1$. It is evident from
Fig.~5 that  $\varepsilon^{\prime}[ \vec{A}^{\text{ext}}]$ has a
maximum as a function of $\beta$ which increases as function of
$L=\Omega^{1/4}$. Moreover the peak shrinks and shifts toward
$\beta=1$ by increasing $L$. Indeed we found the following
pseudocritical couplings: $\beta_c \simeq 0.92 \,, 0.97 \,, 0.99$ for
$L_2=L_3=6, \,8$ and $10$ respectively. 
In order to obtain the critical
parameters for the infinite lattice we should apply the finite size
scaling analysis~\cite{Barber89} to our data. We plan to present the
results of this analysis in a future publication.
Nevertheless, it si gratifying to see that our preliminary results 
corroborate the theoretical expectation that
$\varepsilon[\vec{A}^{\text{ext}}]$ behaves as an energy density.

\section{Conclusions}

We have presented a new method that allows to investigate the
effective action for external background fields in gauge systems by
means of Monte Carlo simulations. Our method has been successfully
tested for the lattice U(1) pure gauge theory. However it can be
extended in a straightforward manner to the non~Abelian gauge
theories.

\clearpage

\section*{FIGURE CAPTIONS}

\renewcommand{\labelenumi}{Figure \arabic{enumi}.}
\begin{enumerate}
\item
The field strength tensor $F_{\mu\nu}$ (Eq.~\eqref{Eq16}) normalized
to the external field (Eq.~\eqref{Eq17}) versus the Euclidean time
$x_4$ on a lattice $64\times10^2\times32$ at $\beta=0.9$ and
$n^{\text{ext}}=2$. Circles, squares, triangles, diamonds, crosses,
and stars refer respectively to the components $(1,2)$, 
$(1,3)$, $(1,4)$, $(2,3)$, $(2,4)$, $(3,4)$ of the field strength
tensor.
\item 
The same quantity as in Fig.~1 at $\beta=1.1$.
\item
The contributions to the derivative of the lattice effective action
density
Eq.~\eqref{Eq20} (in units of $\varepsilon^{\prime}_{\text{ext}}$)
due to the plaquettes in the various $(\mu,\nu)$
planes versus the Euclidean time $x_4$ at $\beta=1.1$ and
$n^{\text{ext}}=2$ on a $64\times10^2\times32$ lattice. The planes are
labeled using the same notations as in Fig.~1.
\item
The derivative of the lattice effective action density
Eq.~\eqref{Eq19} (in units of $\varepsilon^{\prime}_{\text{ext}}$)
versus $\beta$ with $n^{\text{ext}}=2$.
Circles, squares, and triangles refer to $L_2=L_3=6, \, 8, \, 10$
respectively.
\item
Figure 4 near the critical region $\beta \approx 1$.
\end{enumerate}

\newpage
\begin{figure}[t]
\begin{center}
\epsfig{file=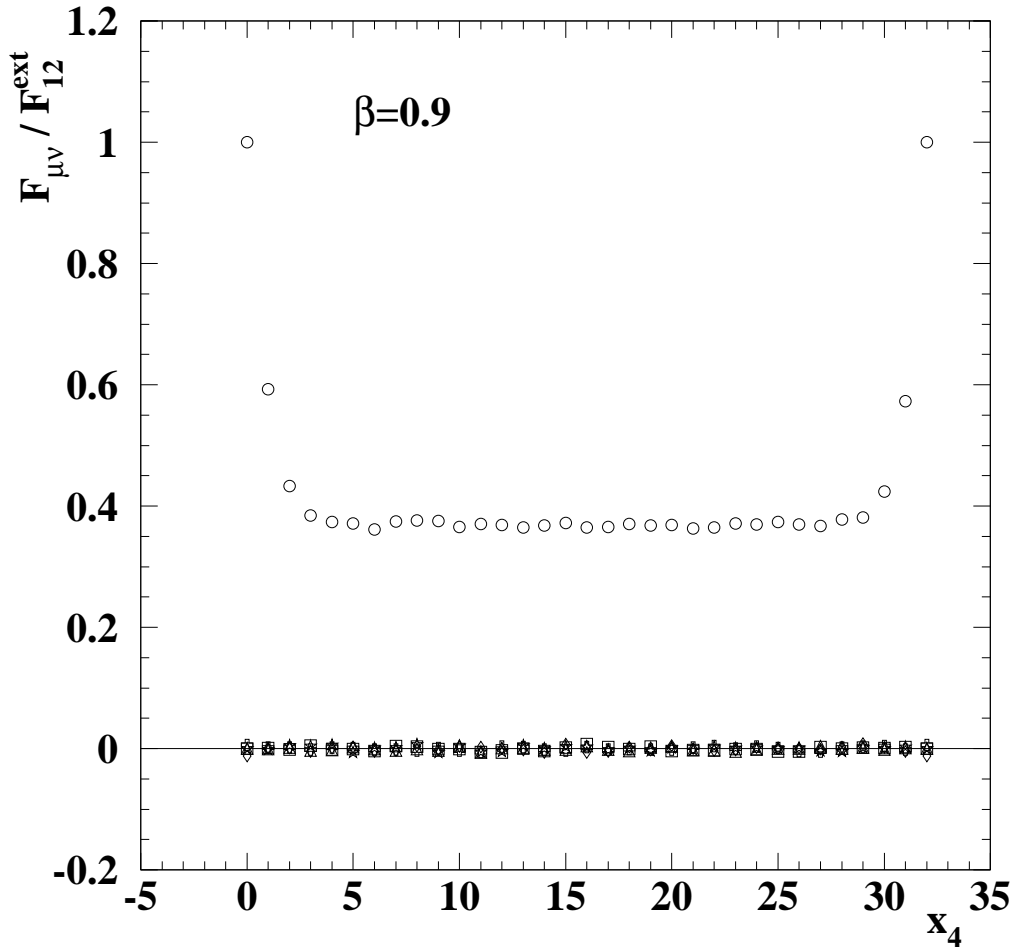,width=\textwidth}
\end{center}
\vspace{-10pt}
\caption{
The field strength tensor $F_{\mu\nu}$ (Eq.~\eqref{Eq16}) normalized
to the external field (Eq.~\eqref{Eq17}) versus the Euclidean time
$x_4$ on a lattice $64\times10^2\times32$ at $\beta=0.9$ and
$n^{\text{ext}}=2$. Circles, squares, triangles, diamonds, crosses,
and stars refer respectively to the components $(1,2)$, 
$(1,3)$, $(1,4)$, $(2,3)$, $(2,4)$, $(3,4)$ of the field strength
tensor.
}
\label{Fig:1}
\end{figure}
\newpage
\begin{figure}[t]
\begin{center}
\epsfig{file=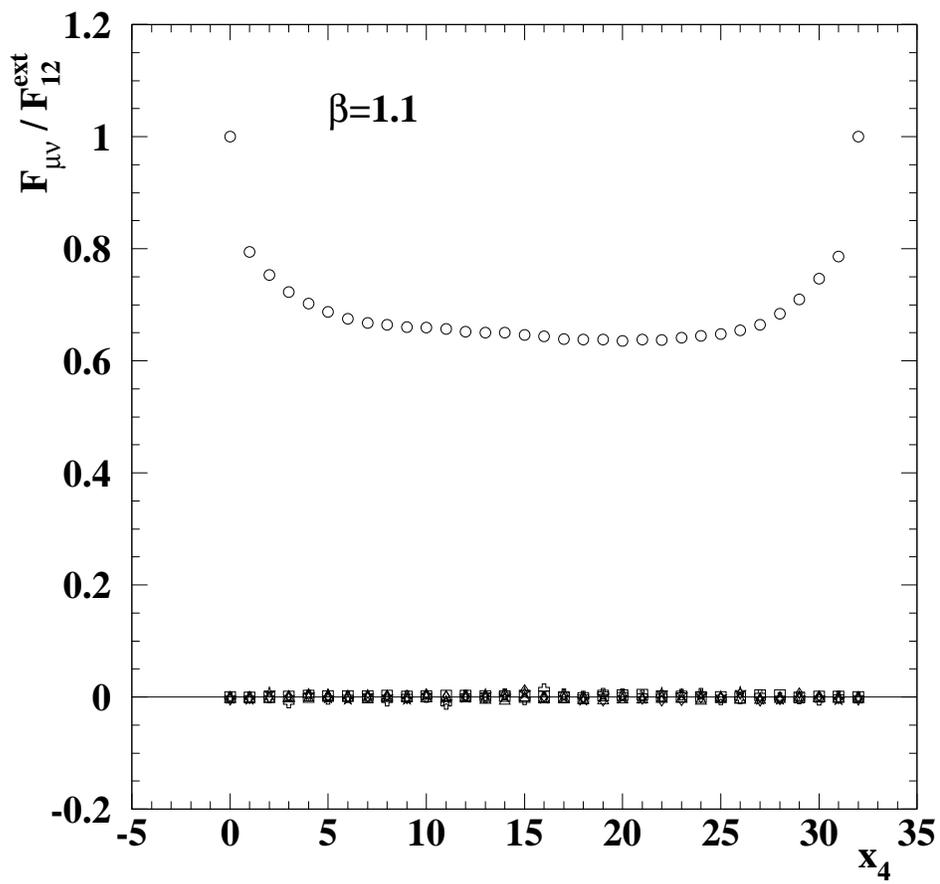,width=\textwidth}
\end{center}
\vspace{-10pt}
\caption{
The same quantity as in Fig.~1 at $\beta=1.1$.
}
\label{Fig:2}
\end{figure}
\newpage
\begin{figure}[t]
\begin{center}
\epsfig{file=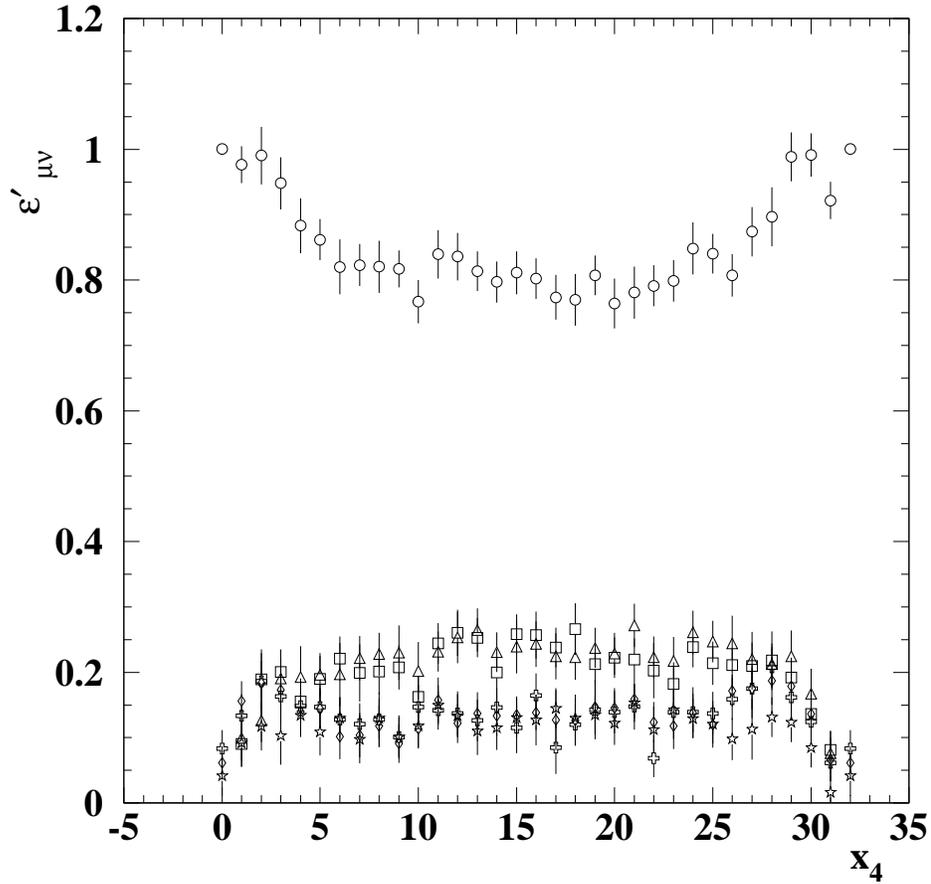,width=\textwidth}
\end{center}
\vspace{-10pt}
\caption{
The contributions to the derivative of the lattice effective action
density
Eq.~\eqref{Eq20} (in units of $\varepsilon^{\prime}_{\text{ext}}$)
due to the plaquettes in the various $(\mu,\nu)$
planes versus the Euclidean time $x_4$ at $\beta=1.1$ and
$n^{\text{ext}}=2$ on a $64\times10^2\times32$ lattice. The planes are
labeled using the same notations as in Fig.~1.
}
\label{Fig:3}
\end{figure}
\newpage
\begin{figure}[t]
\begin{center}
\epsfig{file=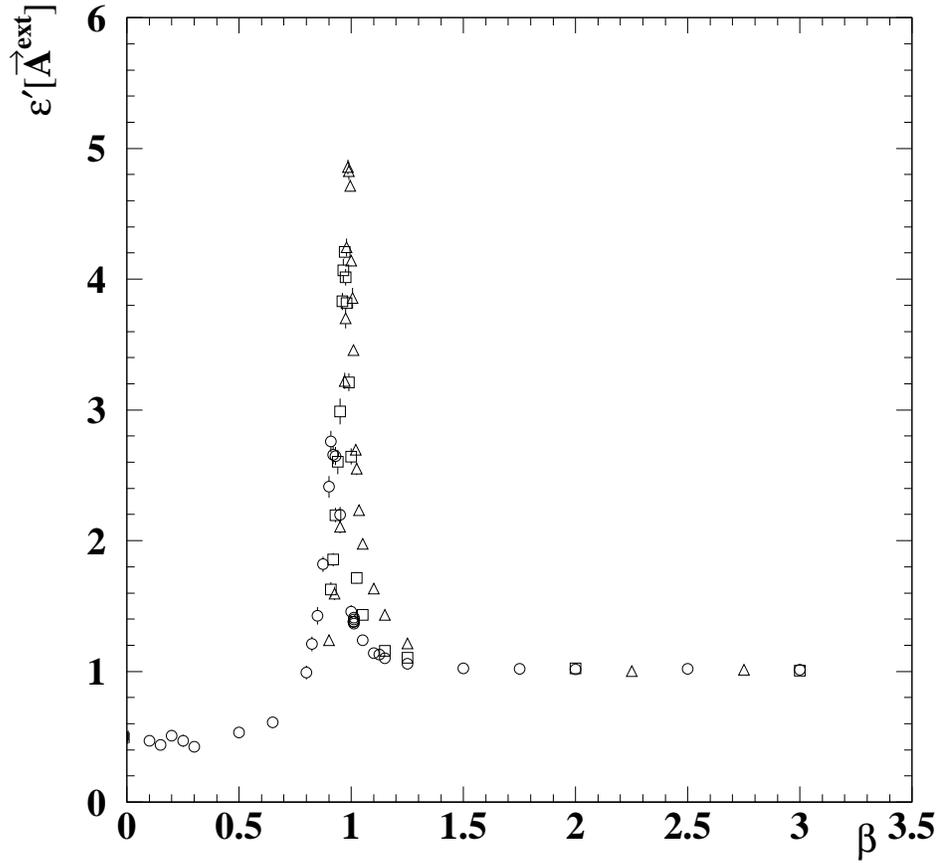,width=\textwidth}
\end{center}
\vspace{-10pt}
\caption{
The derivative of the lattice effective action density
Eq.~\eqref{Eq19} (in units of $\varepsilon^{\prime}_{\text{ext}}$)
versus $\beta$ with $n^{\text{ext}}=2$.
Circles, squares, and triangles refer to $L_2=L_3=6, \, 8, \, 10$
respectively.
}
\label{Fig:4}
\end{figure}
\newpage
\begin{figure}[t]
\begin{center}
\epsfig{file=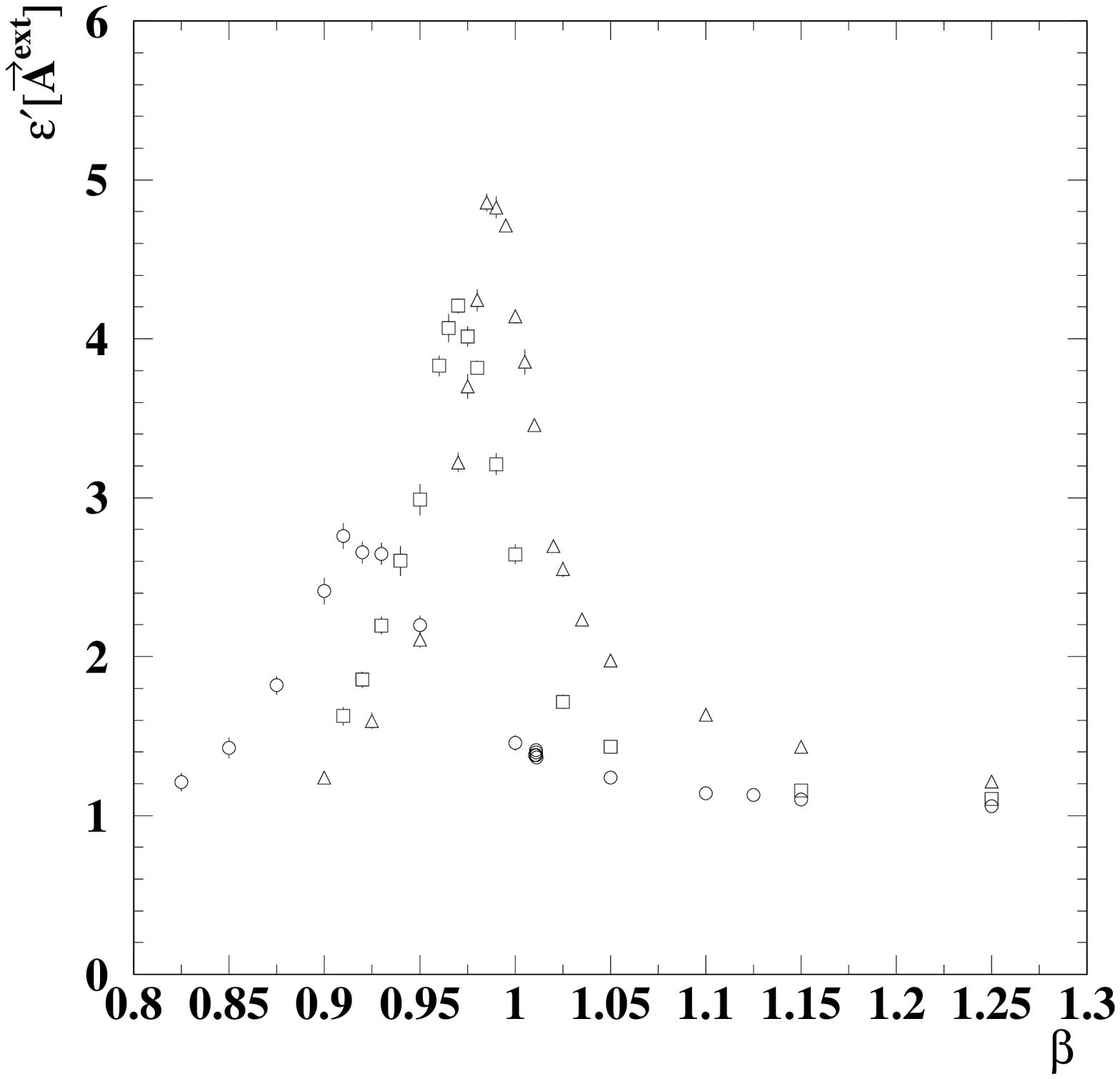,width=\textwidth}
\end{center}
\vspace{-10pt}
\caption{
Figure 4 near the critical region $\beta \approx 1$.
}
\label{Fig:5}
\end{figure}


\begin{thebibliography}{99}
\bibitem{Wolff86} U. Wolff, Nucl. Phys. {\bf B265} (1986) 506; {\em
ibid.} 537. 
\bibitem{Luscher92} M. L\"uscher, R. Narayanan, P. Weisz, and U.
Wolff, Nucl. Phys. {\bf B384} (1992) 168.
\bibitem{Banks77} T. Banks, R. Myerson, and J. Kogut, Nucl. Phys. {\bf
B129} (1977) 493.
\bibitem{DeGrand80} T. A. DeGrand and D. Toussaint, Phys. Rev. {\bf
D22} (1980) 2478.
\bibitem{Kerler95} W. Kerler, C. Rebbi, and A. Weber, Phys. Lett. {\bf
B348} (1995) 560.
\bibitem{DelDebbio95} L. Del Debbio, A. Di Giacomo, and G. Paffuti,
Phys. Lett. {\bf B349} (1995) 513; Nucl. Phys. {\bf B} (Proc. Suppl.)
{\bf 42} (1995) 231.
\bibitem{Hasenfratz90} A. Hasenfratz, P. Hasenfratz, and F.
Niedermayer, Nucl. Phys. {\bf B329} (1990) 739.
\bibitem{Barber89} N. Barber, in {\em Phase Transitions and Critical
Phenomena} Vol.8, C. Domb and J. E. Lebowitz eds., Academic Press,
1989.
\end{thebibliography}
\end{document}